\documentclass[letterpaper, 10 pt, conference]{ieeeconf}  % Comment this line out
\IEEEoverridecommandlockouts                              % This command is only
\usepackage{amssymb}
\usepackage{graphicx}
\graphicspath{{./figure/}}
\usepackage{float}
\usepackage{color}
\usepackage{mathrsfs} 
\usepackage{mathtools}
\usepackage{caption}
\usepackage{subcaption}
\usepackage{bm}
\usepackage{hyperref}
\usepackage{cleveref}

\usepackage{amsthm}

\theoremstyle{plain}
\newtheorem{theorem}{Theorem}

\newtheorem{remark}{Remark}

\theoremstyle{definition}

\newtheorem{problem}{Problem}
\newtheorem{assumption}{Assumption}

\definecolor{cadmiumgreen}{rgb}{0.0, 0.42, 0.24}

\newcommand{\R}{\mathbb{R}}

\newcommand{\Gb}{\bm{G}}
\newcommand{\Qb}{\bm{Q}}

\DeclareMathOperator*{\E}{\mathbb{E}}
\DeclareMathOperator*{\diag}{diag}

\title{\LARGE \bf Youla-REN: Learning Nonlinear Feedback Policies with Robust Stability Guarantees}

\author{Ruigang Wang and Ian R. Manchester% <-this % stops a space
	\thanks{This work was supported by the Australian Research Council.}% <-this % stops a space
	\thanks{The authors are with the Australian Centre for Field Robotics, The University of Sydney, Sydney, NSW 2006, Australia
		(e-mail: {\tt\small  ian.manchester@sydney.edu.au}).}%
}

\begin{document}

\maketitle
\pagestyle{empty}

%%%%%%%%%%%%%%%%%%%%%%%%%%%%%%%%%%%%%%%%%%%%%%%%%%%%%%%%%%%%%%%%%%%%%%%%%%%%%%%%
\begin{abstract}
This paper presents a parameterization of nonlinear controllers for uncertain systems building on a recently developed neural network architecture, called the recurrent equilibrium network (REN), and a nonlinear version of the Youla parameterization. The proposed framework has “built-in” guarantees of stability, i.e., all policies in the search space result in a contracting (globally exponentially stable) closed-loop system. Thus, it requires very mild assumptions on the choice of cost function and the stability property can be generalized to unseen data. Another useful feature of this approach is that policies are parameterized directly without any constraints, which simplifies learning by a broad range of policy-learning methods based on unconstrained optimization (e.g. stochastic gradient descent). We illustrate the proposed approach with a variety of simulation examples.
\end{abstract}

\section{Introduction}

Neural networks have recently gained popularity in various control tasks due to their success in machine learning and artificial intelligence (e.g. \cite{silver2017mastering}). Many existing work focuses on learning neural network controllers for unknown dynamical systems in the framework of reinforcement learning (RL) \cite{sutton2018reinforcement}. Despite the potential of solving hard control problems, there are still well-known issues for the RL controllers, such as sample complexity and interpretability, which impedes their applications in complex nonlinear system with critical safety requirement \cite{amodei2016concrete,brunke2021safe}. 
%Safety and stability guarantee is the central problem in RL-based control learning \cite{berkenkamp2017safe,chow2018lyapunov,fisac2018general,taylor2020learning,brunke2021safe}.

Even for the most classic control design setting where mathematical model of the system is available, it is still a challenge problem of learning provable stabilizing controllers \cite{chang2019neural}. An intuitive way is to parameterize both stability certificate (i.e. Lyapunov function) and control policy via deep neural networks (DNNs), and then use constrained optimization method to ensure that the corresponding Lyapunov inequality holds for the training data \cite{mehrjou2019deep,berkenkamp2017safe,dai2021lyapunov}. The stability property usually depends on the training data and may not be able to generalized to unseen data. Similar approaches have been applied to learn control barrier functions such that the system state remains in a safety set \cite{taylor2020learning}.  

Another approach is to project the neural network parameters into a stabilizing policy set based on classic stability analysis method for models rather than sampled data \cite{gu2021recurrent,kretchmar2001robust}. Although provable stability is guaranteed, this approach often has higher computation cost for large-scale neural networks since the set of stabilizing policies is often highly non-convex. A convex inner approximation was developed in \cite{gu2021recurrent}. 

For RL problems with linear system setting, \cite{roberts2011feedback} demonstrated that the Youla policy parameterization (\cite{youla1976modern}) can guarantee closed-loop stability and offer a number of performance advantages over some natural and naive parameterizations. More historical details about Youla parameterization are referred to \cite{boyd1991linear}, and extensions to nonlinear systems can be found in \cite{van2000l2}.

In this paper, we consider designing robust state-feedback controllers for uncertain linear systems such that the accumulated running cost is also minimized. Although the system is linear, the optimal controller for general cost is nonlinear, e.g., model predictive control is a nonlinear policy with the presence of state/input constraint. Our approach can also be extended to other general system setups, such as nonlinear system and partially observed system.

{\bf Contributions.} The main contribution of this work is a novel parameterization of nonlinear controllers, called \emph{Youla-REN}, which builds on  recently developed neural network architecture, called the recurrent equilibrium network (REN) \cite{revay2021recurrent}, and a nonlinear version of the Youla parameterization. The proposed controller set has ``built-in'' guarantees of stability, that is, all policies in the search space result in a contracting (globally exponentially stable) closed-loop system. Such stability guarantees do not rely on the choice of cost function, the length of rollout trajectories, and training data distribution, which makes it generalizable for unseen data and suitable for various control tasks. Another useful feature of this approach is that all policies are parameterized directly without any constraints, which allows for easy and scalable learning via many unconstrained optimization methods, e.g. stochastic gradient descent (SGD). Finally, we demonstrate the effectiveness of the proposed method via a variety of numerical examples. 

{\bf Paper outline.} Section~\ref{sec:problem} gives a formal problem formulation. Section~\ref{sec:approach} presents the proposed Youal-REN policy parameterization, followed by various simulation examples in Section~\ref{sec:example-1}.

{\bf Notation.} We use $\mathcal{U}(X)$ to denote the uniform distribution over a compact set $X$. We use bold upper letters $ \Gb,\Qb,\ldots$ to represent dynamical systems and bold lower letters $ \bm{x},\bm{y},\ldots $ to denote discrete-time signals, i.e., $\bm{x}=(x_0,x_1,\ldots)$. The rest of the notation is standard.

\section{Problem Formulation}\label{sec:problem}
Consider uncertain linear dynamical systems of the form:
\begin{equation}\label{eq:system}
	x_{t+1}=A(\rho)x_t+Bu_t+w_t
\end{equation}
with measured state $ x_t\in \R^{n}$, input $ u_t\in \R^{m}$, uncertain parameter $\rho \in \mathbb{P}$ and disturbance $ w_t\sim \mathcal{D}(\mathbb{W})$, where $\mathbb{P},\mathbb{W} $ are compact sets, and $\mathcal{D}$ is a known distribution. Our proposed method can be easily extended more general nonlinear and partially observed systems, see discussions in Section~\ref{sec:extension}.
% \begin{equation}
% 	x_{t+1}=f(\rho,x_t)+Bu_t+w_t.
% \end{equation}

The control performance is the average value of a cost over trajectories of length $ T$, i.e.,
\begin{equation}
	\ell(\bm{x},\bm{u},\bm{w})=\frac{1}{T+1}\sum_{t=0}^T c(x_t,u_t)
\end{equation}
where $ \bm{x}=(x_0,x_1,\ldots,x_T),\ \bm{u}=(u_0,u_1,\ldots,u_T),\ \bm{w}=(w_0,w_1,\ldots,w_T)$ are the trajectories of \eqref{eq:system} over the horizon $[0,T]$. The stage cost function $ c$ is assumed to be piecewise differentiable. We wish to design a feedback controller of the form
\begin{equation}\label{eq:policy}
	u=\pi_\theta(x)
\end{equation}
where $\theta\in \Theta\subseteq \R^N$ is trainable parameter, such that it (at least approximately) solves the following problem
\begin{equation}\label{eq:J-theta}
	\min_{\theta\in \Theta} \; J(\theta)=\E_{\substack{\rho\sim \mathcal{U}(\mathbb{P}) \\ w_t\sim \mathcal{D}(\mathbb{W})}} \bigl[\ell(\bm{x},\bm{u},\bm{w})\mid \pi_\theta\bigr].
\end{equation}
For general stage costs, the optimal controller is nonlinear even if the system is linear and certain. 

Since it is generally hard to solve \eqref{eq:J-theta} exactly, an alternative way is to search for an approximate solution using data-driven approaches. That is, starting with an initial guess $\theta^0$, during the $k$th iteration, we first generate $M$ independent scenarios $ (\rho^i, x_0^i, \bm{w}^i)$, where $\rho^i$ is the uncertain parameter, $x_0^i$ is the initial state and $ W^i$ is a disturbance sequence. Then, we compute the empirical cost 
\begin{equation}
	\hat{J}(\theta)=\frac{1}{M}\sum_{i=1}^M \ell(\bm{x}^i,\bm{u}^i,\bm{w}^i).
\end{equation}
where $(\bm{x}^i,\bm{u}^i,\bm{w}^i)$ is the trajectory rollout.
Finally, the control parameter is updated via
\begin{equation}\label{eq:sgd}
	\theta^{k+1}=\theta^k-\alpha^k \nabla \hat{J}(\theta^k)
\end{equation}
where $\alpha^k>0$ is a step size. The basic requirement for the above data-driven approach is that $\pi_\theta $ is a robustly stabilizing controller, which is often achieved by imposing extra stability constraints on $\pi_\theta$. This usually leads to a complex and probably non-convex constraint on $\theta$, and constrained optimization methods are computationally expensive for learning large-scale neural network controllers. To address those issues, this work mainly focuses on the following problem.
 
\begin{problem}\label{prob:1}
	Construct a unconstrained robust policy parameterization (i.e., $\Theta=\R^N$) such that $ x_{t+1}=A(\rho)x_t+B\pi_\theta(x_t)$ is globally exponentially stable for all $\rho\in \mathbb{P}$ and $ \theta\in \R^N$.
\end{problem}

\section{Youla Parameterization via REN}\label{sec:approach}

In this section, we first recall a recently developed neural network architecture -- recurrent equilibrium network \cite{revay2021recurrent}. Then, we use it to construct a nonlinear Youla parameterization for the uncertain linear system. Extensions to more general system settings are also discussed. 

\subsection{Recurrent equilibrium networks}

\begin{figure}[!bt]
    \centering
    \includegraphics[width=0.5\linewidth]{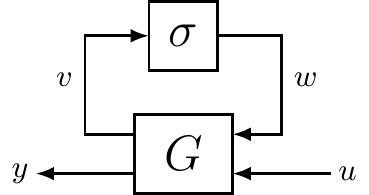}
    \caption{An Lur'e system perspective of REN}\label{fig:lure}
\end{figure}

REN is a nonlinear dynamical system of the form: 
\begin{equation}\label{eq:ren}
    \begin{split}
        \renewcommand\arraystretch{1.1}
        \begin{bmatrix}
            x_{t+1} \\ v_t \\ y_t
        \end{bmatrix}&=
        \overset{W}{\overbrace{
		\left[
            \begin{array}{c|cc}
            A & B_1 & B_2 \\ \hline 
            C_{1} & D_{11} & D_{12} \\
            C_{2} & D_{21} & D_{22}
		\end{array} 
		\right]
        }}
        \begin{bmatrix}
            x_t \\ w_t \\ u_t
        \end{bmatrix}+
        \overset{b}{\overbrace{
            \begin{bmatrix}
                b_x \\ b_v \\ b_y
            \end{bmatrix}
        }}\\
        w_t=\sigma&(v_t):=
        \begin{bmatrix}
            \sigma(v_{t}^1) & \sigma(v_{t}^2) & \cdots & \sigma(v_{t}^q)
        \end{bmatrix}^\top
    \end{split}
\end{equation} 
where $ x_t\in \R^{n_x},u_t \in \R^{n_u},y_t\in \R^{n_y}$ are the state, input and output, respectively. $ v_t,w_t\in \R^{n_v}$ are the input and output of the neuron layer. We assume that the activation function is $ \sigma:\R\rightarrow\R$ with slope restricted in $[0,1]$. In this work, we will use rectified linear unit (ReLU) $\sigma(x)=\max(x,0)$ as the default activation for RENs. The learnable parameter is $\theta'=(W,b)$ where $W$ is the weight matrix and $b$ is the bias vector. 

The REN can also be viewed as an Lur'e system, see Fig.~\ref{fig:lure}. The feedback structure forms an \emph{implicit} or \emph{equilibrium} neuron layer:
\begin{equation}\label{eq:equilibrium}
    w_t=\sigma(D_{11}w_t+C_1x_t+D_{12}u_t+b_v),
\end{equation}
whose solutions are also the equilibrium points of the difference equation $w_t^{k+1}=\sigma(Dw_t^k+b_w)$ or the ordinary differential equation $\frac{d}{ds}w_t(s)=-w_t(s)+\sigma(Dw_t(s)+b_w) $, where $b_w=C_1x_t+D_{12}u_t+b_v$ is ``frozen'' for each time-step. Solving \eqref{eq:equilibrium} online is equivalent to running an infinite depth feedforward network \cite{bai2019deep}. The matrix $D_{11}$ can be interpreted as the adjacency matrix of the graph defining interconnections between the neurons. By imposing different block structure on $D_{11}$, we can divide the implicit layer into many sub-layers and formulate complex network topology, including DNN, CNN and ResNet, etc \cite{el2021implicit}.

Since the nonlinear activation function is slope-restricted in $[0,1]$, the neuron layer $\sigma$ satisfies the following incremental integral quadratic constraints (IQCs):
\begin{equation}
    \begin{bmatrix}
        \Delta v_t \\ \Delta w_t
    \end{bmatrix}^\top
    \begin{bmatrix}
        0 & \Lambda \\
        \Lambda & -2\Lambda
    \end{bmatrix}
    \begin{bmatrix}
        \Delta v_t \\ \Delta w_t
    \end{bmatrix}\geq 0,\quad \forall t\in \mathbb{N}
\end{equation}
where $ \Lambda\in \R^{n_v\times n_v}$ is a positive diagonal matrix, $ (\Delta \bm{v},\Delta \bm{w})$ is the difference between any pair of input-output trajectories of $\sigma$. From IQC theorem \cite{Megretski:1997}, we can conclude that the REN satisfies the incremental IQC defined by $(Q,S,R)$: 
\begin{equation}
    \sum_{t=0}^{T}\begin{bmatrix}
        \Delta y_t \\ \Delta u_t
    \end{bmatrix}^\top
    \begin{bmatrix}
        Q & S^\top \\
        S & R
    \end{bmatrix}
    \begin{bmatrix}
        \Delta y_t \\ \Delta u_t
    \end{bmatrix}\geq 0,\quad \forall T\in \mathbb{N}
\end{equation}
where $ 0\succeq Q \in \R^{n_y\times n_y}$, $S\in \R^{n_u\times n_y}$ and $ 0 \preceq R \in \R^{n_u\times n_u}$, if there exists a positive-definite $ P\in \R^{n_x\times n_x}$ and a positive diagonal matrix $\Lambda\in \R^{n_v\times n_v}$ such that 
\begin{equation}\label{eq:lmi-qsr-explicit}
    \begin{split}
        \begin{bmatrix}
            P & -C_1^\top \Lambda & C_2^\top S^\top\\
            -\Lambda C_1 & W & D_{21}^\top S^\top - \Lambda D_{12} \\
            S C_2 & S D_{21} - D_{12}^\top \Lambda & R +SD_{22}+D_{22}^\top S^\top
        \end{bmatrix}- \\
        \begin{bmatrix}
            A^\top \\ B_1^\top \\ B_2^\top 
        \end{bmatrix}P
        \begin{bmatrix}
            A^\top \\ B_1^\top \\ B_2^\top 
        \end{bmatrix}^\top -
        \begin{bmatrix}
            C_2^\top \\ D_{21}^\top \\ D_{22}^\top
        \end{bmatrix}
        Q
        \begin{bmatrix}
            C_2^\top \\ D_{21}^\top \\ D_{22}^\top
        \end{bmatrix}^\top \succ 0.
    \end{split}
\end{equation}
Important special cases of incremental IQCs include:
\begin{itemize}
    \item $Q=-\frac{1}{\gamma}I, S=0, R=\gamma I$: the REN satisfies an $\ell_2$ Lipschitz bound, a.k.a. incremental $\ell^2$-gain bound, of $\gamma$.
    \item $Q=0,S=I, R=0$: the REN satisfies incremental passivity condition.
\end{itemize}
An intuitive way to learn RENs is through constrained optimization. However, LMI \eqref{eq:lmi-qsr-explicit} quickly becomes the computational bottleneck as the model size increases. A central result of \cite{revay2021recurrent} is a \emph{direct parameterization} of all well-posed RENs. Roughly speaking, by applying certain transform $\theta'=f(\theta)$, Condition \eqref{eq:lmi-qsr-explicit} is automatically satisfied for any $\theta\in \R^N$. That is, learning an REN becomes an unconstrained optimization problem under the new coordinate $\theta$.

Throughout the rest of this paper, we will utilize a subclass of REN, called \emph{acyclic} REN (aREN) where the weight $D_{11}$ is constrained to be strictly lower triangular. That is, the equilibrium layer \eqref{eq:equilibrium} has feedforward structure. The major benefit of aREN is its simple implementation as \eqref{eq:equilibrium} yields an explicit solution. Various learning tasks in \cite{revay2021recurrent} shows that aREN often provides similar quality of models as REN. 

%It is also worth to point out that aREN contains all stable linear systems by simply setting the weighting matrices except $A,B_2,C_2,D_{22}$ to zero.

\subsection{Youla-REN}

First, we make the following assumption on the uncertain linear system \eqref{eq:system}. 
\begin{assumption}\label{asmp:1}
    There exists a robust controller of the form:
    \begin{equation}\label{eq:K}
        u=-Kx+v
    \end{equation}
    where $v\in\R^m$ is an additional control augmentation, such that system \eqref{eq:system} has a finite $\ell_2$-gain bound from $v$ to $x$.
\end{assumption}
This is equivalent to that the uncertain linear system \eqref{eq:system} is robustly stabilizable by linear state-feedback control. With the extra control augmentation $v$, we are able to search for a policy with better control performance while maintaining the robust stability guarantee. The basic idea is to build on a standard method for \emph{linear} feedback optimization: the Youla-Kucera parameterization, a.k.a Q-augmentation \cite{youla1976modern,zhou1996robust}. 

\begin{figure}[!bt]
    \centering
    \begin{tabular}{cc}
        \includegraphics[width=0.42\linewidth]{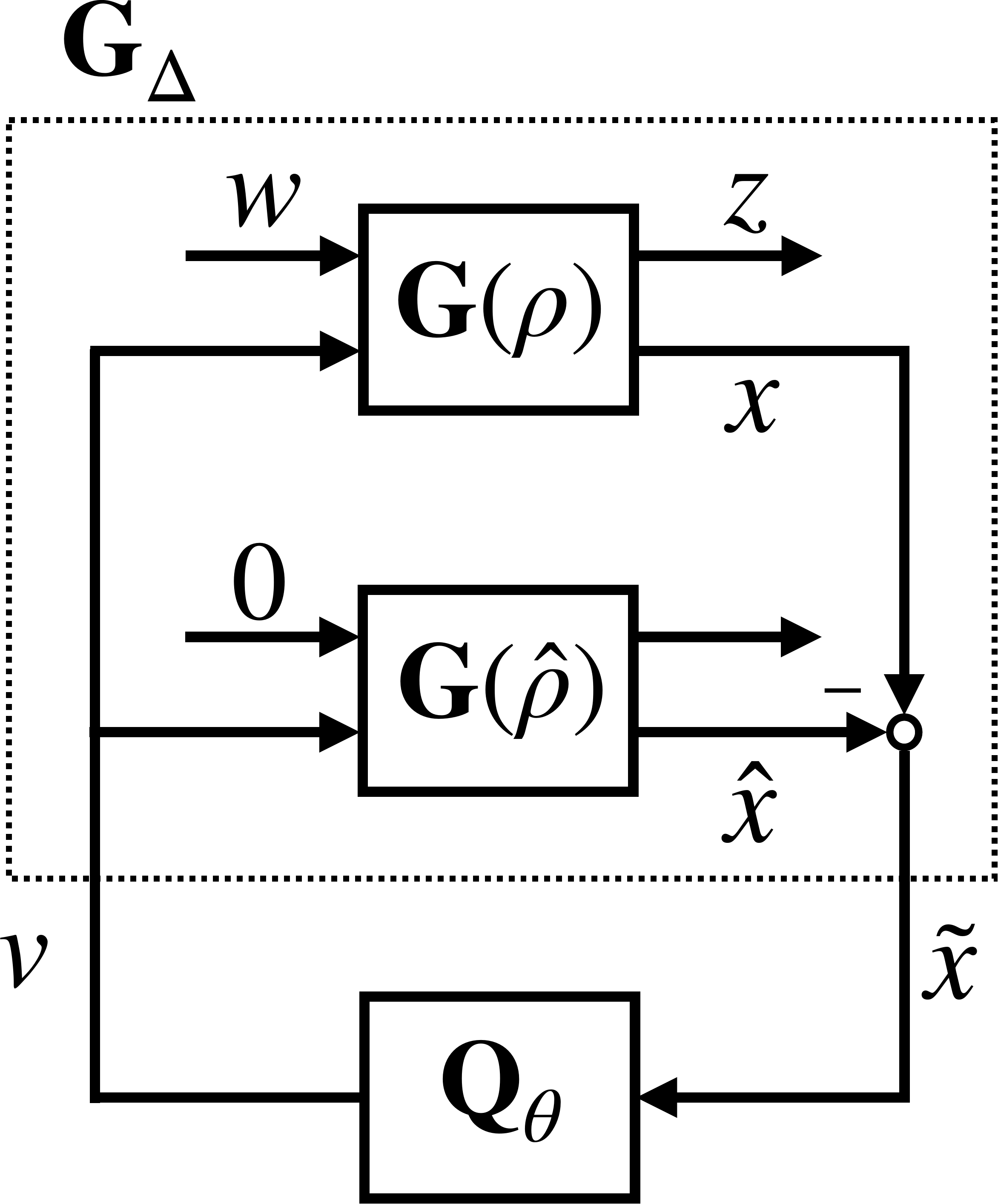} &
        \includegraphics[width=0.45\linewidth]{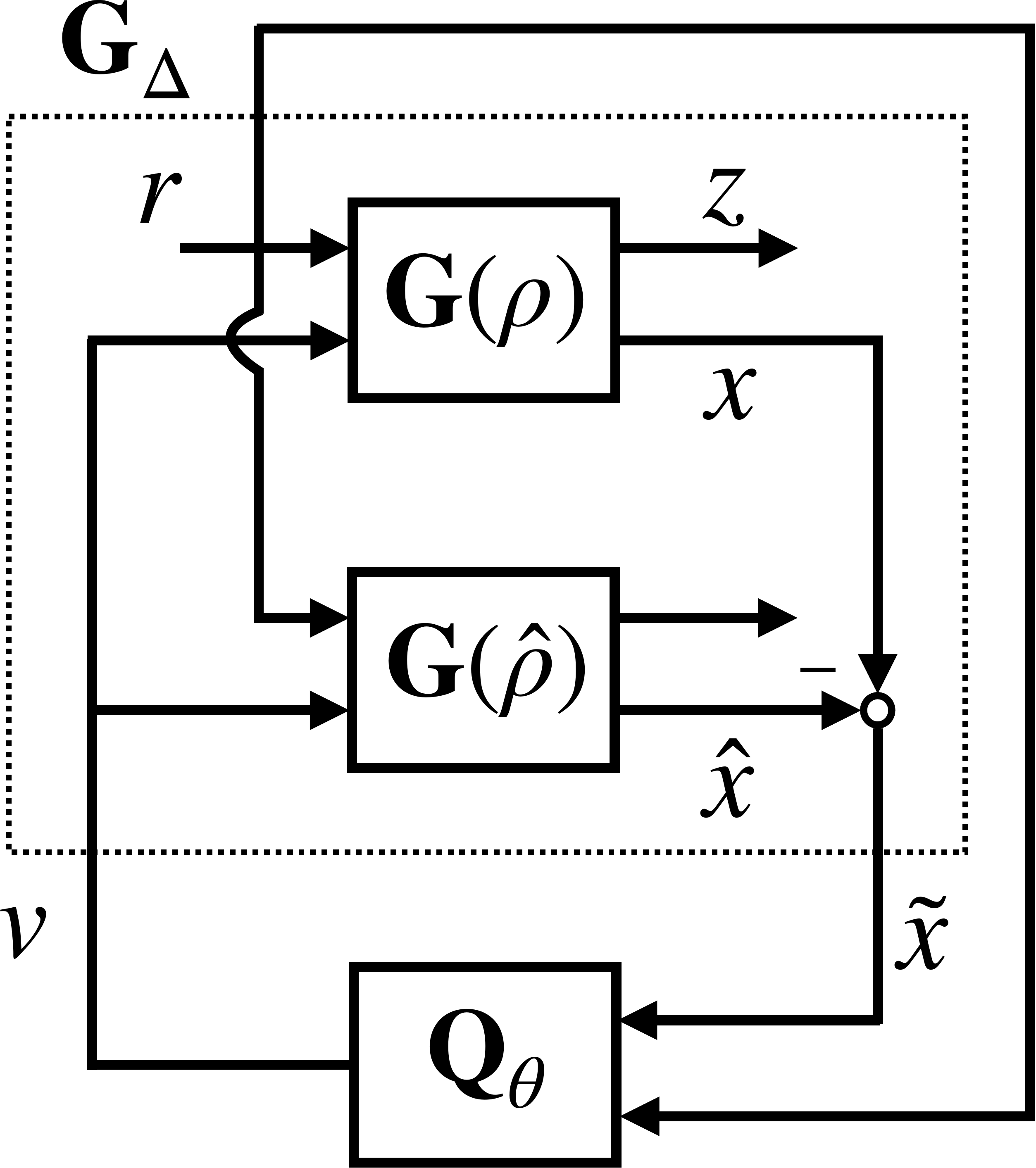} \\
        (a) disturbance rejection & (b) reference tracking
    \end{tabular}
    \caption{Youla policy parameterization}\label{fig:youla}
\end{figure}

Letting $ z=(x,u,w)$ be the performance output, the closed-loop dynamics can be written as the transfer matrix
\begin{equation}
    \begin{bmatrix}
        \bm{x} \\ \bm{z}
    \end{bmatrix}=
    \overset{\Gb(\rho)}{\overbrace{
    \begin{bmatrix}
        \Gb_{xv}(\rho) & \Gb_{xw}(\rho) \\
        \Gb_{zv}(\rho) & \Gb_{zw}(\rho)
    \end{bmatrix}}}
    \begin{bmatrix}
        \bm{v} \\ \bm{w}
    \end{bmatrix}
\end{equation}
where $ \Gb_{xv},\Gb_{xw},\Gb_{zv},\Gb_{zw}$ are stable for all $\rho \in \mathbb{P}$. Now let us consider the scheme plot in Fig.~\ref{fig:youla}a, where the dynamics of $\Gb_\Delta$ can be described by
\begin{equation}
    \begin{bmatrix}
        \tilde{\bm{x}} \\ \bm{z}
    \end{bmatrix}=
    \begin{bmatrix}
        \Gb_{xv}(\rho)-\Gb_{xv}(\hat{\rho}) & \Gb_{xw}(\rho) \\
        \Gb_{zv}(\rho) & \Gb_{zw}(\rho)
    \end{bmatrix}
    \begin{bmatrix}
        \bm{v} \\ \bm{w}
    \end{bmatrix}
    % \tilde{\bm{x}}= \Gb_\Delta(\rho,\hat{\rho})\bm{v} :=\bigl[\Gb_{xv}(\rho)-\Gb_{xv}(\hat{\rho})\bigr]\bm{v} 
\end{equation}
where $\hat{\rho}$ is a nominal value chosen from $\mathbb{P}$. Note that the above system is incrementally stable.

% It is well-known that if $\rho=\hat{\rho}$, the set of all stabilizing linear feedback controllers can be parameterized by stable linear systems $ \Qb:\tilde{\bm{x}}\mapsto \bm{v}$. We can further enlarge the searching space from stable linear $\Qb$ to incrementally stable nonlinear $\Qb$. 

\begin{theorem}\label{thm:main}
    Suppose that Assumption~\ref{asmp:1} holds for system \eqref{eq:system} and $\Gb_\Delta$ admits the incremental IQC defined by 
    \begin{equation*}
        \overset{Q}{\overbrace{\begin{bmatrix}
            Q_{xx} & 0 \\
            0 & Q_{zz}
        \end{bmatrix}}}\prec 0, \quad
        \overset{S}{\overbrace{\begin{bmatrix}
            S_{vx} & 0 \\
            0 & S_{wz}
        \end{bmatrix}}},\quad 
        \overset{R}{\overbrace{\begin{bmatrix}
            R_{vv} & 0 \\
            0 & R_{ww}
        \end{bmatrix}}}\succ 0.
    \end{equation*} 
    Let $\Qb_\theta$ with $\theta\in \R^N$ be an REN satisfying the incremental IQC defined by $ (\overline{Q},\overline{S},\overline{R})$. Then, the closed-loop system is contracting and yields finite Lipschitz bound from $ w$ to $z$ for all $\rho \in \mathbb{P}$ if 
    \begin{equation}\label{eq:stability}
        \begin{bmatrix}
            Q_{xx}+\overline{R} & S_{vx}^\top + \overline{S} \\
            S_{vx}+\overline{S}^\top & R_{vv}+\overline{Q}
        \end{bmatrix}\prec 0.
    \end{equation}
\end{theorem}
\begin{remark}
    If $ Q_{xx}=-\frac{1}{\alpha}I, S_{vx}=0, R_{vv}=\alpha I$ with $\alpha>0$, i.e., $\Gb_\Delta$ has incremental $\ell_2$-gain bound of $\alpha$ from $ v$ to $x$, then Condition~\eqref{eq:stability} can be reduced to the requirement for incremental small-gain theorem. That is, we can make the closed-loop system contracting by choosing $\Qb_\theta$ with $\overline{Q}=-\frac{1}{\gamma}I, \overline{S}=0, \overline{R}=\gamma I$ for some positive constant $\gamma < 1/\alpha$. For the reference tracking problem, we can feed the reference $\bm{r}$ to the Youla parameter $\Qb_\theta$, as shown in Fig.~\ref{fig:youla}b, and specify an arbitrarily large but finite gain bound for $\bm{r}$ to $\bm{v}$, that is, $ \overline{Q} = -\frac{1}{\gamma}I$ and $\overline{R}=\diag(\gamma I, \eta I)$ where $\eta\gg \gamma$, which can help $\Qb_\theta$ learn the mapping between reference and nominal input.
\end{remark}
We call the following controller an \emph{Youla-REN} policy: 
\begin{equation}\label{eq:youla}
    \pi_\theta:
    \begin{cases}
        \hat{x}_{t+1} = [A(\hat{\rho})-BK]\hat{x}_t+Bv_t \\
        v_t=\Qb_\theta(x_t-\hat{x}_t) \\
        u_t=-Kx_t+v_t
    \end{cases}
\end{equation}
where $\hat{x}_t$ is the state of $\Gb(\hat{\rho})$. By introducing nonlinearity in $\Qb_\theta$, we can significantly increase the expressive power of the candidate policy set, which is useful for learning optimal policy subject to general cost functions.

\subsection{Extensions to more complex systems} \label{sec:extension}
\subsubsection{Nonlinear systems}
For certain class of continuous-time nonlinear systems, there exist several constructive methods for designing controllers that render the closed-loop system contracting \cite{manchester2017control}, virtually contracting \cite{wang2021nonlinear} and robustly contracting \cite{manchester2018robust}. The proposed Youla-REN can be naturally integrated with those methods by introducing an augmented control input, which is similar to \cite{van2000l2}.

\subsubsection{Partially observed systems}
When only partial information $y=Cx$ is available for \eqref{eq:system}, we can construct a standard output-feedback structure with $v_t$ as additional control augmentation \cite{boyd1991linear}: 
\begin{align}
    \hat{x}_{t+1}&=A(\hat{\rho})+Bu_k+L\tilde{y}_t \label{eq:observer}\\
    \tilde{y}_t&=y_t-C\hat{x}_t \\
    u_t &= -K\hat{x}_t+v_t
\end{align}
where the observer gain $L$ is designed such that \eqref{eq:observer} is robustly stable. By estimating the incremental $\ell_2$-gain bound for $\Gb_\Delta: \bm{v}\mapsto \tilde{\bm{y}}$, we can construct robustly stabilizing policy set via RENs $\Qb_\theta:\tilde{\bm{y}}\mapsto \bm{v}$. For partially observed nonlinear systems, \cite{yi2021reduced} developed constructive methods for building globally converging observers based on contraction analysis.  
%\subsection{Decentralized/distributed systems}

\section{Examples} \label{sec:example-1}

In this section, we will illustrate the proposed approach via a variety of numerical simulations. 

\subsection{System setup for linearized cart-pole system}
Let $p_x$ be the cart position and $\psi$ be the angular displacement of the pendulum from its vertical position. The control task is to balance an inverted pendulum resting on top of a cart (i.e. $ \psi=0$ ) by exerting horizontal forces $u\in \R$ on the cart. For a pendulum of length $\ell$ and mass $M_p$, and for a cart of mass $ M_c$, the linearized dynamics of the cart-pole system at the vertical position are 
\begin{equation}\label{eq:cp-ldyn}
    \begin{split}
        \dot{x}=&
    \begin{bmatrix}
        0 & 1 & 0 & 0 \\
        0 & 0 & -\frac{M_pg}{M_c} & 0 \\
        0 & 0 & 0 & 1 \\
        0 & 0 & \frac{(M_c+M_p)g}{M_c\ell} & 0
    \end{bmatrix}x+
    \begin{bmatrix}
        0 \\ \frac{1}{M_c} \\ 0 \\ -\frac{1}{M_c}
    \end{bmatrix}u \\
    :=& A(\rho)x+Bu
    \end{split}
\end{equation}
where $ x=(p_x,\dot{p}_x,\psi,\dot{\psi})$ and $ \rho=M_p$ are the state and uncertain parameter, respectively. Model parameters are given by $M_p\in [0.2,2]$, $ M_c=1$, $l=1$ and $ g=9.81 $. 

We design a robust controller \eqref{eq:K} by solving the following parametric LMIs:
\begin{equation}\label{eq:lmi}
    \begin{split}
        \min_{\alpha,X,Y}\; & \quad \beta \\
        \mathrm{s.t.}\; &
        \begin{bmatrix}
        W(\rho) & B & X \\
        B^\top & -\beta I & 0 \\
        X & 0 & -\beta I
        \end{bmatrix} \preceq 0, \\
        & W(\rho) + 2\lambda X \succeq 0,\ X \succeq 0
    \end{split}
\end{equation}
where $ W(\rho)= XA^\top(\rho)+A(\rho)X+BY+Y^\top B^\top$. The first LMI implies that the closed-loop system achieves $\mathcal{L}_2$-gain bound of $\beta$ from $v$ to $x$. By minimizing $\beta$, we wish to have a small $\mathcal{L}_2$-gain bound $\alpha$ for $\Gb_\Delta$ and a large set of $Q_\theta$ by Thm.~\ref{thm:main}. The second LMI in \eqref{eq:lmi} means that the convergence rate of closed-loop system is smaller than $\lambda$, avoiding aggressive gain $K$. By choosing $\lambda=5$, we obtain a robust controller \eqref{eq:K} with
\[
    K=YX^{-1}=\begin{bmatrix}
        -7.40 & -14.96 & -125.82 & -27.73
    \end{bmatrix}
\]
and the corresponding gain bound for $Q_\theta$ can be estimated as $\gamma=60$ via Thm.~\ref{thm:main}. Finally, we discretize the linearized cart-pole system \eqref{eq:cp-ldyn} with sampling time $t_s=0.05$. 

\subsection{REN vs RNN/LSTM}\label{sec:Q-learn}

We first consider a quadratic regulation problem with cost 
\begin{equation}\label{eq:qr}
    c(x,u)=x^\top Q x + R u^2
\end{equation}
where $Q=\diag(10,0.1,10,0.1)$ and $ R=0.01$. We will compare the performance of Youla control policy \eqref{eq:youla} with the following four choices of $\Qb_\theta$:  REN, long short-term memory (LSTM) \cite{hochreiter1997long} and vanilla recurrent neural network (RNN) \cite{elman1990finding} with ReLU and tanh activations, referred as RNNr and RNNt, respectively. 

% The RNN can be represented by a  state-space model of the form 
% \begin{equation}\label{eq:rnn}
%     \begin{split}
%         h_{t+1}&=\sigma_h(W_h h_t + U u_t +b_h) \\
%         y_t &= W_y h_t +b_y
%     \end{split}
% \end{equation} 
% where $h_t\in \R^{n_h}, u_t\in \R^{n_u}, y_t\in \R^{n_y}$ are the hidden state, input and output, respectively. We will investigate two activation functions - ReLU $\sigma_h(x)=\max(x,0)$ and hyperbolic tangent $ \sigma_h(x)=\tanh(x)$ where the corresponding models are referred as RNN-ReLU and RNN-tanh, respectively.

% The LSTM contains four components (the input gate $i_t$, output gate $o_t$, forget gate $ f_t$ and memory cell $c_t$, all of them have the same dimension $n_h$) and its dynamics have the following state-space representation
% \begin{equation}\label{eq:lstm}
%     \begin{split}
%         f_{t+1}&=\sigma_s(W_f f_t + U_f u_t + b_f) \\
%         i_{t+1}&=\sigma_s(W_i i_t + U_i u_t + b_i) \\
%         o_{t+1}&=\sigma_s(W_o o_t + U_o u_t + b_o) \\
%         \tilde{c}_{t+1}&=\tanh(W_c c_t + U_c u_t + b_c) \\
%         c_{t+1}&=f_{t+1}\circ c_t + i_{t+1}\circ \tilde{c}_{t+1} \\
%         h_{t+1}&=o_{t+1}\circ \tanh(c_{t+1}) \\
%         y_t & = W_y h_t + b_y
%     \end{split}
% \end{equation}
% where the operator $\circ$ denotes the Hadamard product (element-wise product) and $\sigma_s$ denotes the sigmoid function. 

\paragraph{Training details} All Youla parameters have approximately $ 255,000$ parameters. That is, the RNN has 500 neurons, the LSTM has 250 neurons and REN has $n_x=40$ states and $ n_v=500 $ neurons. We train those policies for 600 epochs with initial learning rate $10^{-3}$ and reduced rate $10^{-4}$ after 400 epochs. During each epoch, it firstly takes $M=10$ random samples of system setups with uniform distribution, i.e., $ (\rho,x_0) \sim \mathcal{U}(\mathbb{P}\times \mathbb{X})$ where $ \mathbb{P}=[0.2,2]$ and $ \mathbb{X}=[-10,10]\times [-0.5,0.5]\times [-2,2] \times [-0.5,0.5]$, then compute $(\hat{J},\nabla \hat{J}) $ based on the closed-loop responses of \eqref{eq:youla} and \eqref{eq:cp-ldyn} over the horizon of $ T=60 $, and finally update the parameter $\theta$ via Adam \cite{Kingma:2017}. Test cost is calculated with $M=50$ system setups and a horizon of $ T = 100$.

\begin{figure}[!bt]
    \centering
    \includegraphics[width=\linewidth]{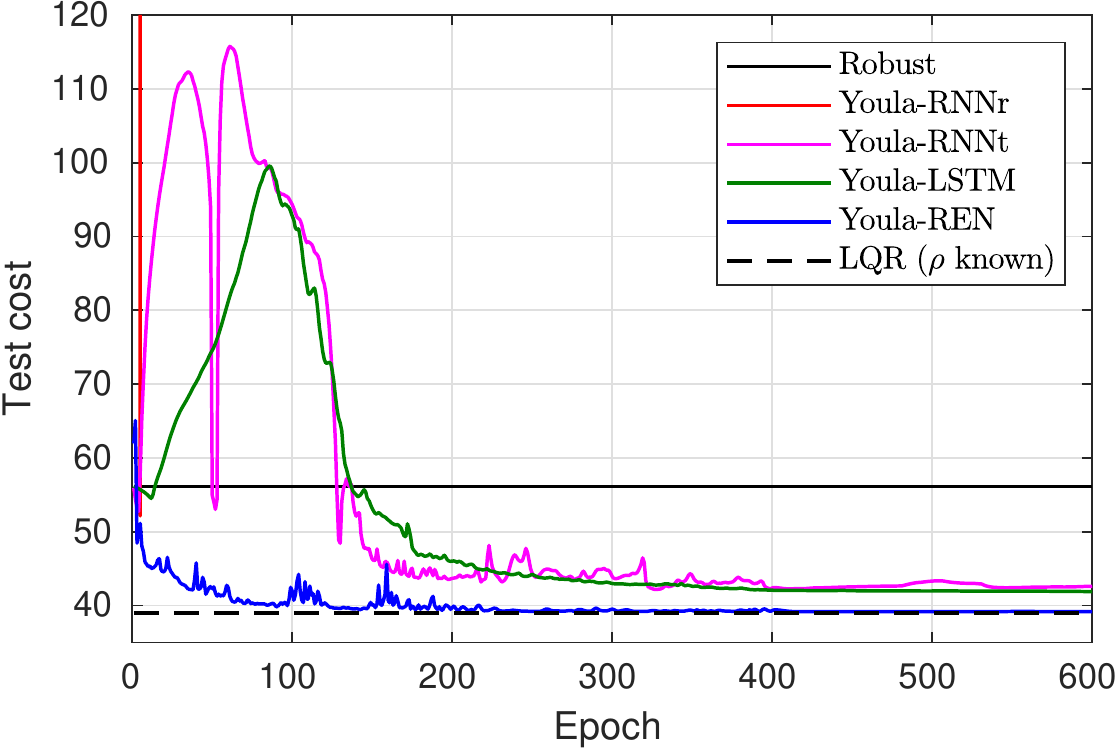}
    \caption{Test cost versus epochs for the linearized cart-pole system.}\label{fig:Q-train}
\end{figure}

\begin{figure}[!bt]
    \centering
    \includegraphics[width=\linewidth]{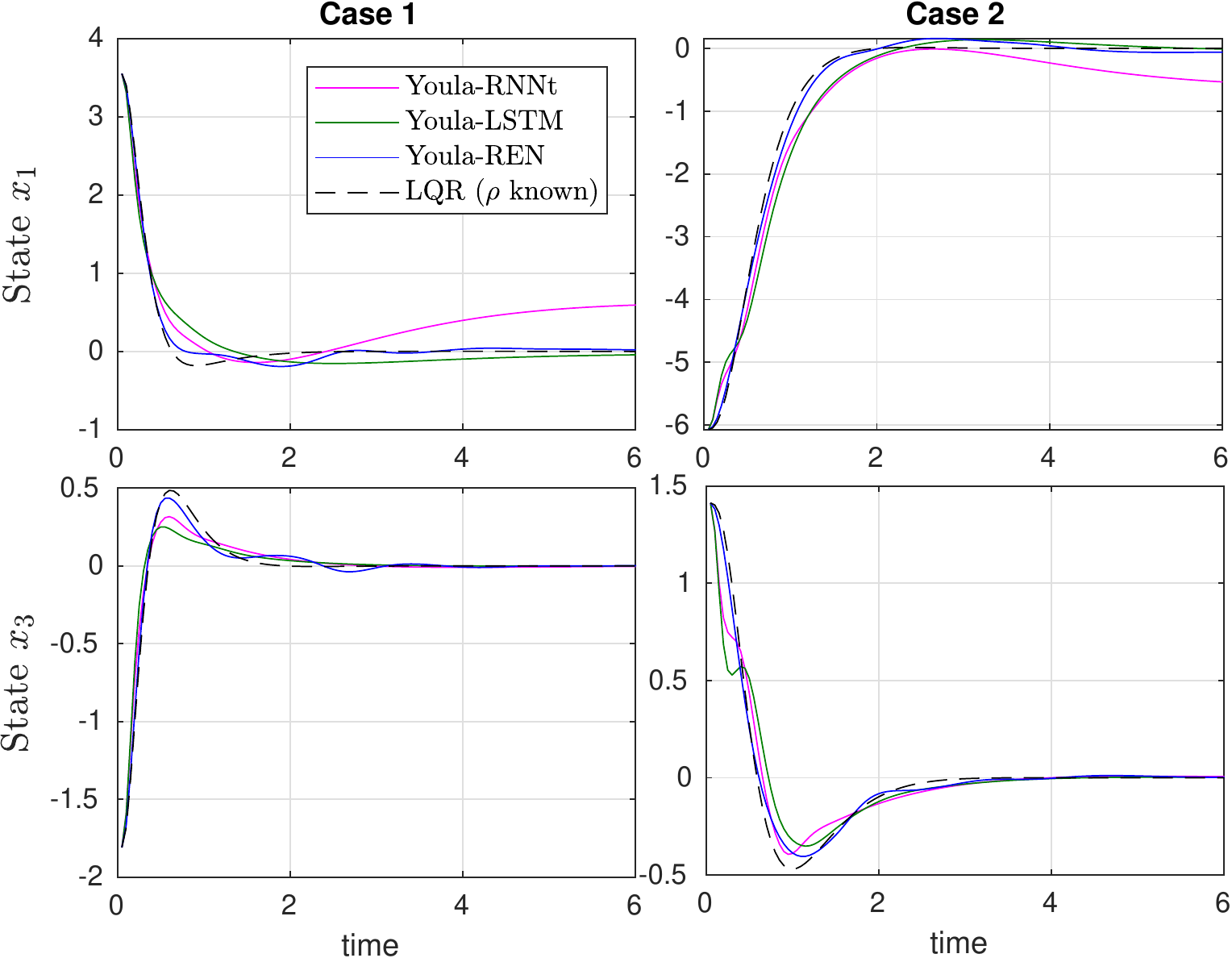}
    \caption{State responses of different Youla policy parameterizations, where each case has different uncertain parameter and initial condition. }\label{fig:Q-sample}
\end{figure}

\paragraph{Results and discussion} We have plotted the test cost versus epochs in Fig.~\ref{fig:Q-train}. The black solid line shows the performance of robust linear controller \eqref{eq:K} with zero augmented input (i.e., $ v=0$), which can be taken as an upper bound of the optimal policy. The black dashed line reveals the performance of optimal LQR controllers with known uncertain parameter $\rho$, which serves as the lower bound of the optimal policy. Firstly, we observed that unstable control policy is found for Youla-RNNr since there is no $\ell_2$-gain regularization applied to $\Qb_\theta$ and ReLU activation is unbounded. Then, we observed that both Youla-RNNt and Youla-LSTM have stable responses although the cost grows significantly larger than the robust linear controller at the first 100 epochs. This is mainly due to the fact that the states of LSTM and RNNt live in some compact sets. But neither Youla-LSTM nor Youla-RNNr can guarantee global exponential stability as their Lipschitz bounds may grow larger than $\gamma$. This can be verified in Fig.~\ref{fig:Q-sample} where the Youla-RNNr yields multiple equilibrium points. After training, the Youla-LSTM and Youla-RNNr have performance gaps $ (\hat{J}-\hat{J}_{opt})/\hat{J}_{opt}$ of $7.72\%$ and $ 9.31\%$, respectively. Thanks to the prescribed Lipschitz bound, Youla-REN can ensure global exponential stability of the closed-loop system. Fig.~\ref{fig:Q-train} shows that the nominal cost decrease quickly and reaches $0.55\%$ performance gap after 250 epochs, which significantly outperforms the other $\Qb$-parameterizations. 

\begin{figure}[!bt]
    \centering
    \includegraphics[width=\linewidth]{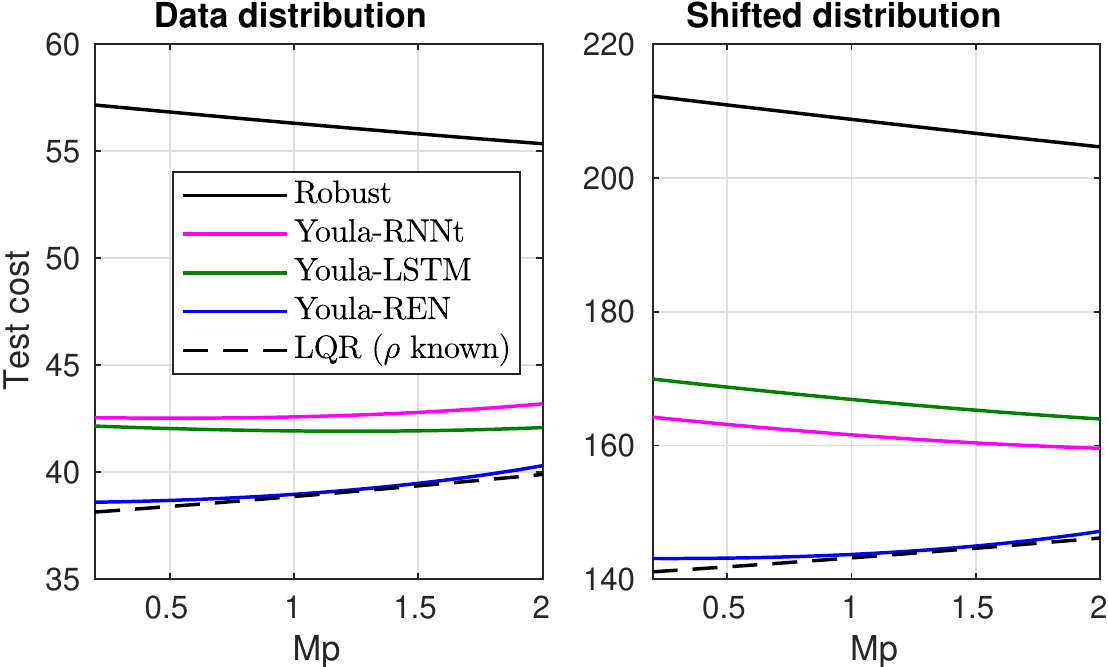}
    \caption{Test cost with training data distribution and unseen distribution versus uncertain parameters. Here the unseen data is from uniform distribution over a shifted initial state set by moving the center of $\mathbb{X}$ from the origin to $ (10, 0, 0, 0) $.}\label{fig:Q-test}
\end{figure}

We have also plotted the test cost versus uncertain parameters for both training data distribution and unseen data distribution in Fig.~\ref{fig:Q-test}. The Youla-REN achieves near optimal performance for the training data and also generalizes to the unseen data. The performance gaps of Youla-RNNt and Youla-LSTM increase to $12.55\%$ and $16.14\%$, respectively, for the unseen data.

\subsection{Youla vs natural control parameterization}
\begin{figure}[!bt]
    \centering
    \includegraphics[width=0.35\linewidth]{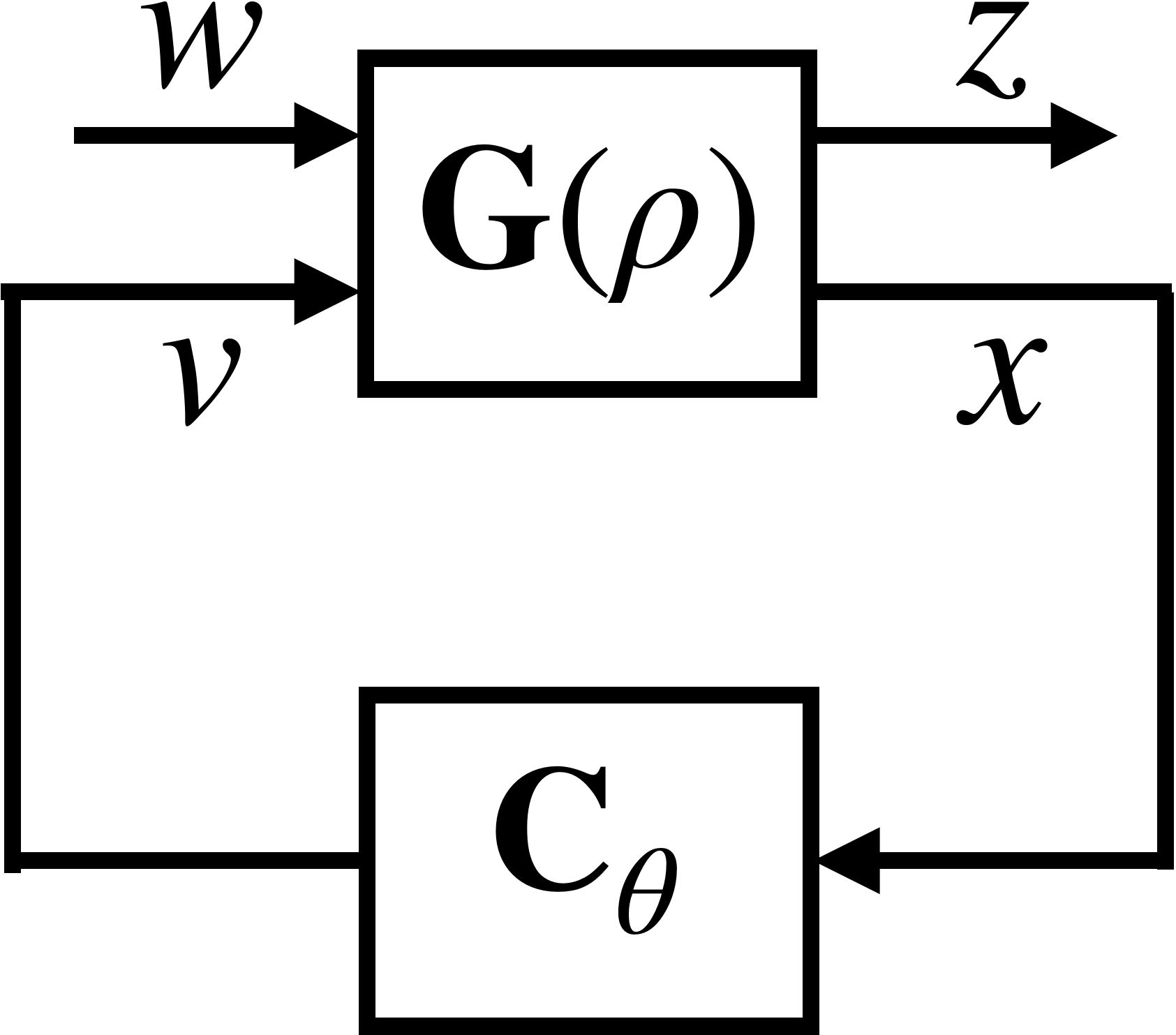} 
    \caption{Natural control policy parameterization}\label{fig:ctrl}
\end{figure}

\begin{figure}[!bt]
    \centering
    \includegraphics[width=\linewidth]{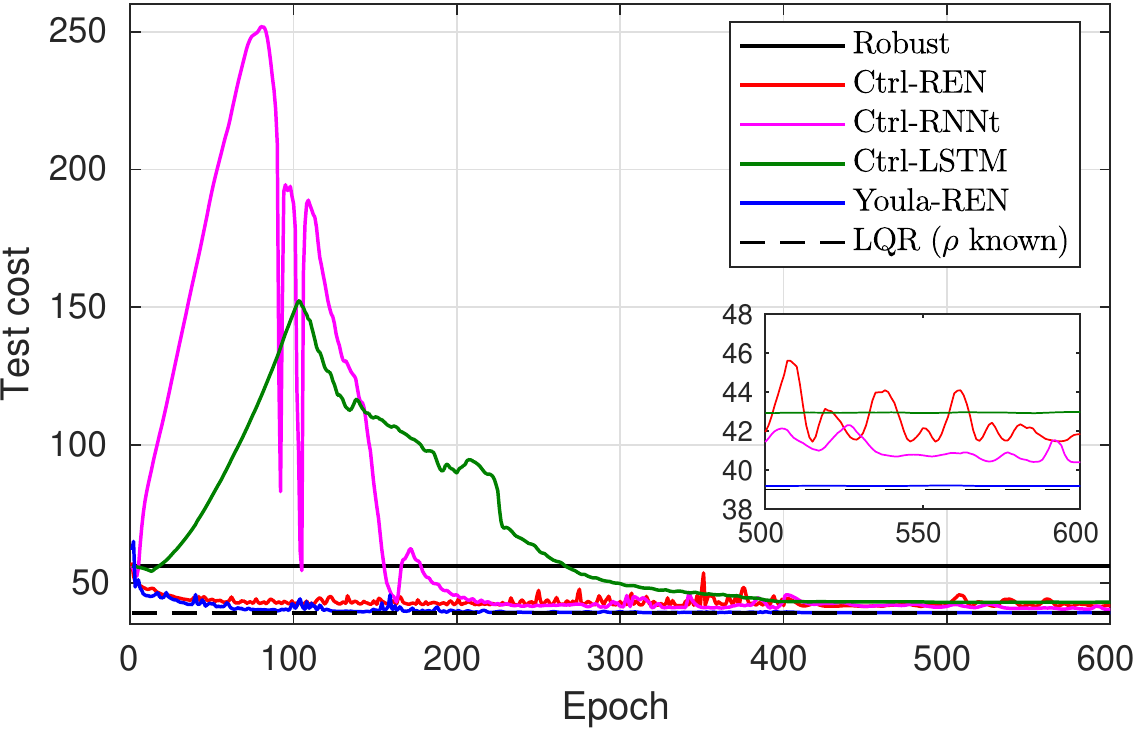}
    \caption{Test cost versus epochs for the quadratic regulation problem using Youla and Ctrl parameterizations.}\label{fig:C-train}
\end{figure}

We here compare the proposed Youla policies with natural control policies (see Fig.~\ref{fig:ctrl}) with $ \bm{v}=\bm{C}_\theta(\bm{x})$, which are denoted by Ctrl-REN, Ctrl-RNNt and Ctrl-LSTM depending on the model used for $\bm{C}_\theta$. We choose the $\ell_2$-gain bound of $\bm{C}_\theta$ to be smaller than $ 1/\beta$ where $\beta$ is obtained by \eqref{eq:lmi}, which ensures closed-loop contracting behavior via incremental small-gain theorem. By comparing the test cost in Fig.~\ref{fig:Q-train} and \ref{fig:C-train}, we observed that Ctrl-LSTM and Ctrl-RNNt have almost doubled peak test cost and also require doubled epochs to learn a controller that outperforms the robust linear policy. Although the policies learned from the Ctrl-REN parameterization generally have decreasing test cost except local spikes, their performance are still worse than Youla-REN. One potential reason is that the gain bound for $\bm{C}_\theta$ in Ctrl-REN is about 6, which is 10 times smaller than $\Qb_\theta$ in Youla-REN since the gain bound of $\Gb$ is usually much larger than $\Gb_\Delta$.  

\subsection{Non-linear vs linear $Q$-parameter}

We now consider the quadratic regulation problem with soft input constraint:
\begin{equation}
    c(x,u)=x^\top Qx+Ru^2+\eta \max(|u|-\overline{u},0)
\end{equation}
with $ \overline{u}=5$ as the bound and $ \eta=50$ as the weighting coefficient. The purpose is to learn a controller which generates control signals $ |u_t|\leq \overline{u}$ when the state is sufficiently close to the set-point. But it is still able to use $|u_t|> \overline{u}$ in a short window to stabilize the system when the state is far way. 

We train both linear and nonlinear $\Qb$ using RENs with $n_x=50, n_v=0$ and $n_x=50, n_v=400$, respectively. For the remaining examples, the initial state set is changed to 
\[
    \mathbb{X}=[-5,5]\times [-0.1,0.1]\times [-1,1] \times [-0.1,0.1].
\] 
The learning procedure is similar to Section~\ref{sec:Q-learn} except it uses $T=100$ and $ T=120 $ for training and testing, respectively.

\begin{figure}[!bt]
    \centering
    \includegraphics[width=\linewidth]{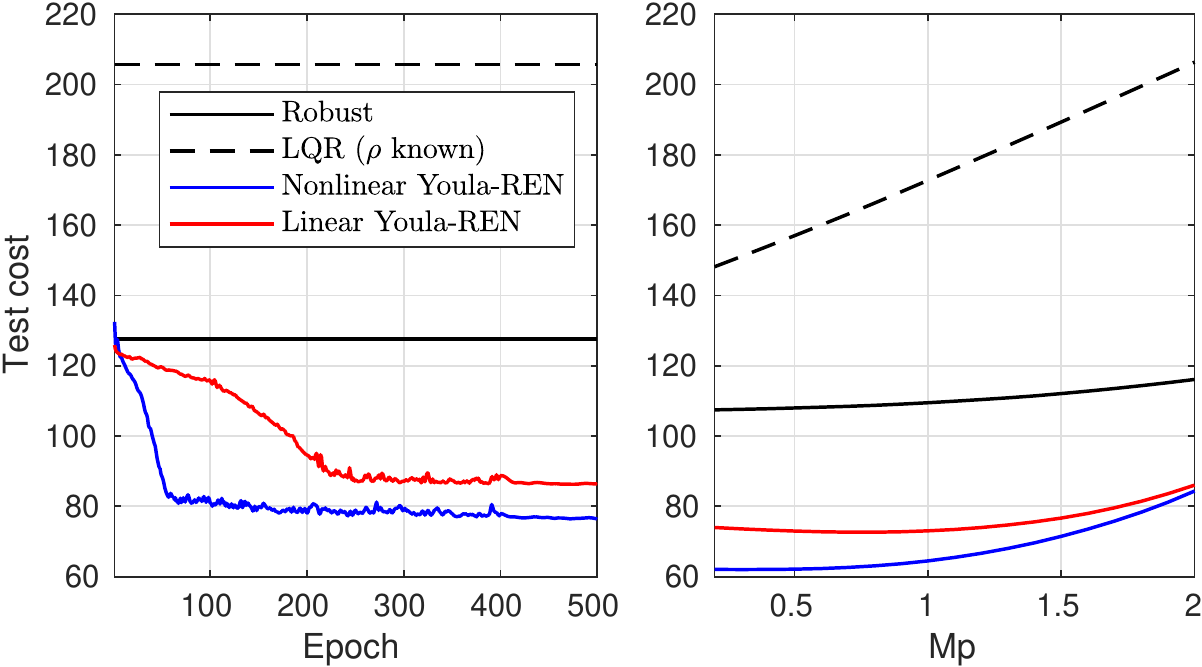}
    \caption{Test cost versus epochs (left) and uncertain parameters (right) for the quadratic regulation problem with soft input constraint.}\label{fig:ub-train}
\end{figure}

\begin{figure}[!bt]
    \centering
    \includegraphics[width=0.9\linewidth]{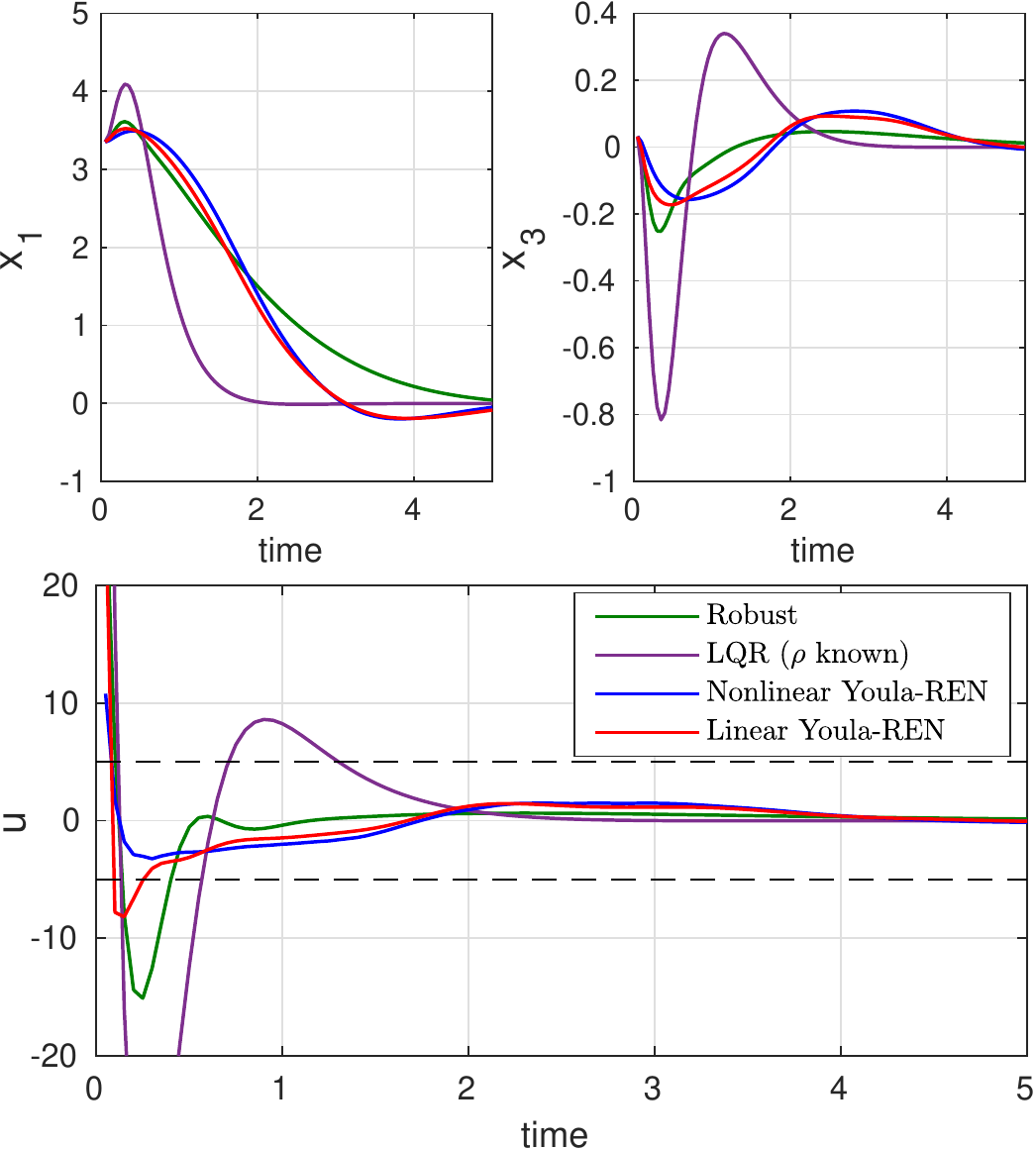}
    \caption{Responses of linear and nonlinear $\Qb$-parameters, where the dashed line indicates the upper and lower bounds.}\label{fig:ub-sample}
\end{figure}

As shown in Fig.~\ref{fig:ub-train}, the nominal cost of nonlinear $\Qb$ decreases faster than the linear one. After 500 epochs, the nonlinear $\Qb$ also outperforms the linear $\Qb$ by 11.5\%. We also have plotted the closed-loop responses in Fig.~\ref{fig:ub-sample}. Their state responses are very close to each other. The main difference is from the input trajectories. When the state is far from the origin, both linear and nonlinear $\Qb$ produce large control actions ($|u|>\overline{u}$). After a few steps, the control signal generated by the nonlinear $\Qb$ remains in the desired range while the linear $\Qb$ still produces excessive input actions.

\subsection{Disturbance rejection}
We revisit the quadratic regulation problem \eqref{eq:qr} but the system is perturbed by unknown input disturbance, i.e.,
\begin{equation}
    \dot{x}=A(\rho)x+B(u+w).
\end{equation}
If $w$ is Gaussian noise and $\rho$ is known, then the optimal policy is LQR controller. For general disturbance types, the optimal controller may be nonlinear. Thus, it is natural to search for a better controller using Youla policy parameterization. Here we consider two scenarios: constant and sinusoidal $ \bm{w}$. For the constant disturbance, we set $ w_t=w_0$ with $t\leq T$ where $ w_0\sim \mathcal{U}([-5,5])$. For the sinusoidal input, we choose $ w_t=A\sin(\omega t+\phi)$ where $ A\sim \mathcal{U}([0,10])$, $ \omega\sim \mathcal{U}([0.05\pi,0.5\pi])$ and $\phi \sim \mathcal{U}([-\pi/2,\pi/2])$.
 
\begin{figure}[!bt]
    \centering
    \begin{tabular}{c}
        \includegraphics[width=0.76\linewidth]{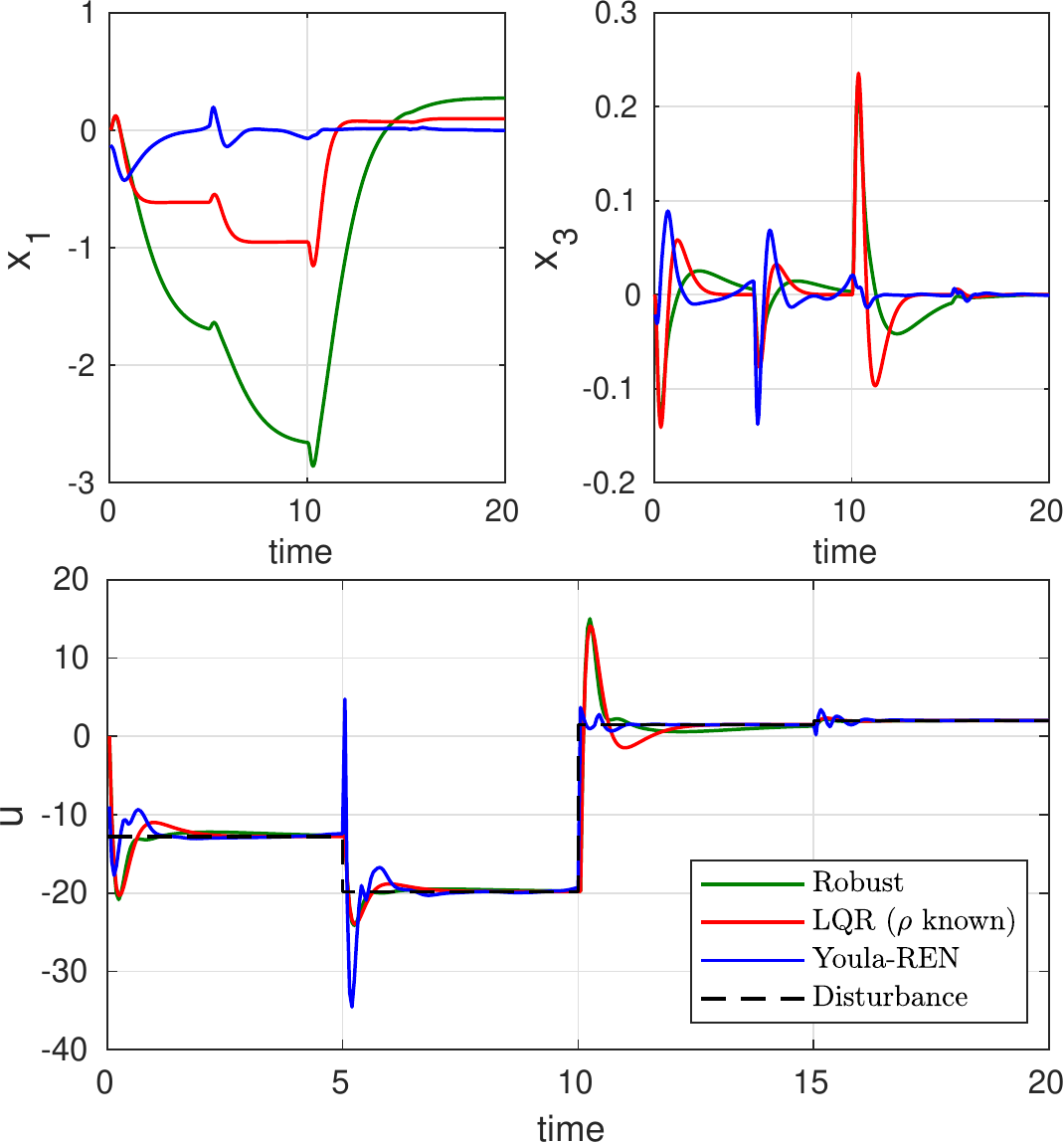} \\
        (a) constant $\bm{w}$ \\
        \includegraphics[width=0.8\linewidth]{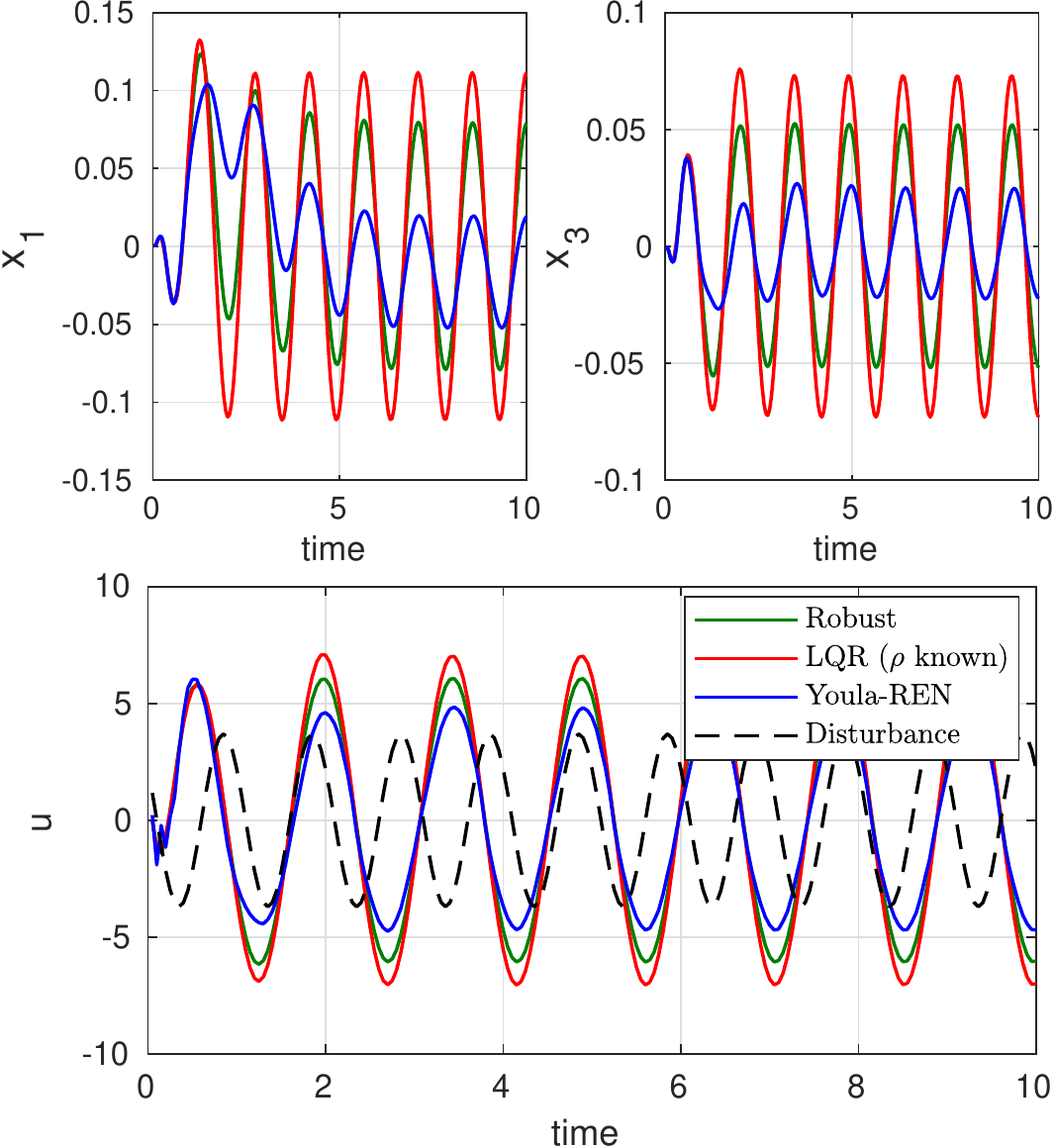} \\
        (b) sinusoidal $ \bm{w}$
    \end{tabular}
    \caption{Responses of Youla-REN to external disturbance.}\label{fig:const-sample}
\end{figure}

Fig.~\ref{fig:const-sample} shows that both robust and LQR controllers relies on steady-state error to cancel the constant disturbance. The Youla-REN can compensate the input disturbance while maintaining the state around the desired equilibrium. For the sinusoidal disturbance, the Youla-REN has smaller amplitude of oscillations in both state and control input.

\subsection{Non-quadratic cost}
\begin{figure}[!bt]
    \centering
    \begin{tabular}{c}
        \includegraphics[width=0.8\linewidth]{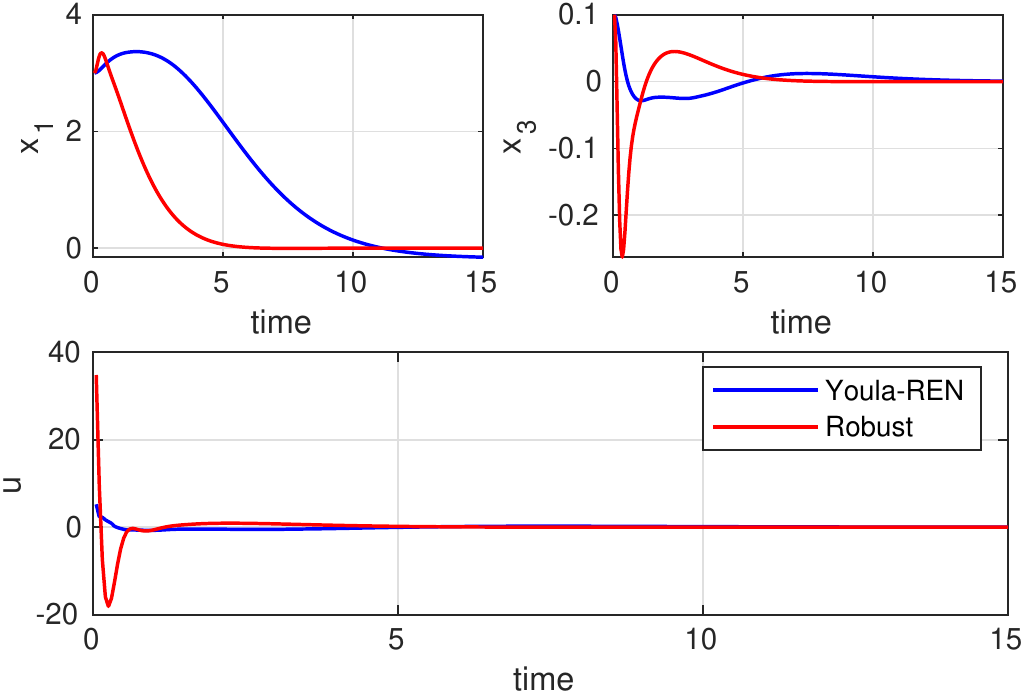} \\
        (a) economic cost \\
        \includegraphics[width=0.8\linewidth]{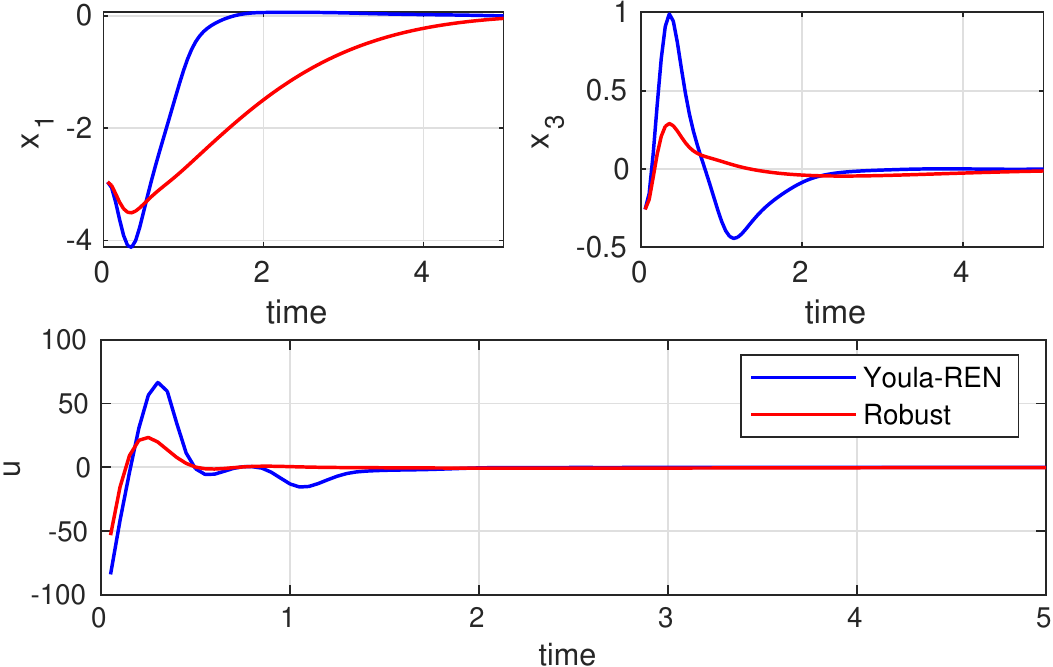} \\
        (b) $\ell_1$ cost
    \end{tabular}
    \caption{Responses of Youla-REN for non-quadratic costs.}\label{fig:non-quadatic}
\end{figure}

We apply the proposed approach to the problems with non-quadratic cost. The first example is the economic cost $ c(x,u)=u^2 $. Note that the optimal policy is simply $ u=0$, which is an unstable controller. By searching in the Youla-REN parameterization, we are able to find a robust stabilizing controller that uses a small amount of control action to stabilize the system, where the state may slowly converge to some non-zero equilibrium point, as shown in Fig.~\ref{fig:non-quadatic}a. The second case is weighted $\ell_1$ cost $c(x,u)=\|W_1x\|_1+\|W_2u\|_1$ where $ W_1=\begin{bmatrix}
    20 & 0.1 & 5 & 0.1
\end{bmatrix}$ and $W_2=0.5$. The result closed-loop response shows that both state and input quickly converge to zero.

\section{Conclusions}
In this work, we have presented a novel control policy parameterization called Youla-REN, which has built-in stability guarantee for uncertain systems. The control policy is flexible and admits a direct parameterization, allowing learning via unconstrained optimization. We have illustrated the benefits of the new policy class via several simulation examples. Our future work will further explore uses of Youla-RENs for robust reinforcement learning and online control.

\appendix
\subsection{Proof of Theorem~\ref{thm:main}}
From the IQC condition for $\Gb_\Delta$, there exists an incremental storage function $ V_g(x^g,\Delta x^g)\geq 0$ with $ V_g(x^g,0)=0$ and $ x^g=(x,\hat{x})$ such that
\begin{equation}
    \begin{split}
        V_g(x_{t+1}^g,&\Delta x_{t+1}^g)-V_g(x_{t}^g,\Delta x_{t}^g)\leq \\
        &\left[
        \begin{array}{c}
        \Delta \tilde{x}_t \\ \Delta z_t \\ \hline \Delta v_t \\ \Delta w_t 
    \end{array} 
    \right]^\top
    \begin{bmatrix}
        Q & S^\top \\
        S & R
    \end{bmatrix}
    \left[
        \begin{array}{c}
        \Delta \tilde{x}_t \\ \Delta z_t \\ \hline \Delta v_t \\ \Delta w_t 
    \end{array} 
    \right].
    \end{split}
\end{equation} 
Similarly, we can also find some incremental storage function $ V_q(x^q,\Delta x^q)\geq 0$ with $ V_q(x^q,0)=0$ where $x^q$ denotes the state of $\Qb_\theta$ such that 
\begin{equation}
    \begin{split}
        V_q(x_{t+1}^q,\Delta &x_{t+1}^q) - V_q(x_{t}^q,\Delta x_{t}^q)\leq \\ &
    \begin{bmatrix}
        \Delta v_t \\ \Delta \tilde{x}_t
    \end{bmatrix}^\top
    \begin{bmatrix}
        \overline{Q} & \overline{S}^\top \\
        \overline{S} & \overline{R}
    \end{bmatrix}
    \begin{bmatrix}
        \Delta v_t \\ \Delta \tilde{x}_t
    \end{bmatrix}.
    \end{split}
\end{equation}
By adding the above inequalities, we have 
\begin{equation}
    \begin{split}
        V_c(&X_{t+1},\Delta X_{t+1})-V_c(X_t,\Delta X_t) \leq \\ &
        \begin{bmatrix}
            \Delta z_t \\ \Delta w_t
        \end{bmatrix}^\top
        \begin{bmatrix}
            Q_{zz} & S_{wz}^\top  \\
            S_{wz} & R_{ww}
        \end{bmatrix}
        \begin{bmatrix}
            \Delta z_t \\ \Delta w_t
        \end{bmatrix}
         + \\ &
        \begin{bmatrix}
            \Delta \tilde{x}_t \\ \Delta v_t
        \end{bmatrix}^\top
        \begin{bmatrix}
            Q_{xx}+\overline{R} & S_{vx}^\top + \overline{S} \\
            S_{vx}+\overline{S}^\top & R_{vv}+\overline{Q}
        \end{bmatrix}
        \begin{bmatrix}
            \Delta \tilde{x}_t \\ \Delta v_t
        \end{bmatrix} \\
        \leq & \begin{bmatrix}
            \Delta z_t \\ \Delta w_t
        \end{bmatrix}^\top
        \begin{bmatrix}
            Q_{zz} & S_{wz}^\top  \\
            S_{wz} & R_{ww}
        \end{bmatrix}
        \begin{bmatrix}
            \Delta z_t \\ \Delta w_t
        \end{bmatrix}
    \end{split}
\end{equation}
where $ V_c=V_g+V_q$ and $ X=(x^g,x^q)$. Here the second inequality follows by Condition~\eqref{eq:stability}. From the above inequality, we can conclude that the closed-loop system is contracting and yields finite Lipschitz bound from $ w$ to $z$.

\bibliographystyle{IEEEtran}
\bibliography{ref}

\end{document}